\documentstyle[aps,prb,epsfig,graphics]{revtex}

\newcommand{\beq}{\begin{equation}}
\newcommand{\eeq}{\end{equation}}
\newcommand{\beqa}{\begin{eqnarray}}
\newcommand{\eeqa}{\end{eqnarray}}
\newcommand{\eps}{\epsilon}
\newcommand{\om}{\omega}
\newcommand{\Om}{\Omega}

\begin{document}
\draft
\title{A theory of the subgap photoemission in one-dimensional electron-phonon
systems. An instanton approach to pseudogaps.}
\author{S. I. Matveenko$^{1,2}$ \footnote{E-mail: matveen@landau.ac.ru} 
 and S. A. Brazovskii$^{2,1}$\footnote{E-mail: brazov@ipno.in2p3.fr}}
\address{$^{1}$ L.D. Landau Institute for Theoretical Physics, Kosygina Str. 2,\\
117940, Moscow, Russia.\\
$^{2}$Laboratoire de Physique Th\'{e}orique et des Mod\`{e}le Statistiques,\\
CNRS, B\^{a}t.100, Universit\'{e} Paris-Sud, 91405 Orsay, France.}
\date{December 25, 2001}
\maketitle

\begin{abstract}
For a one-dimensional electron-phonon system we consider the photon
absorption involving electronic excitations within the pseudogap energy
range. Within the adiabatic approximation for the electron - phonon
interactions these processes are described by nonlinear configurations of an
instanton type. We calculate intensities of the photoelectron spectroscopy
PES including the momentum resolved one ARPES and supplement to known
results for the optical subgap absorption. We start with the generic case of
a one dimensional semiconductor with pronounced polaronic effect. In details
we consider the Peierls model for a half-filled band of electrons coupled to
the lattice which describes the polyacethylene and some commensurate Charge
Density Waves. Particular attention was required for studies of momentum
dependencies for the ARPES where we face an intriguing interference between
the time evolution and the translational motion of the instantons.
\end{abstract}


\section{Introduction.}

This article is devoted to theory of pseudogaps in electronic spectra as
they can be observed by means of the Photo Electron Spectrography (PES) or
the Angle Resolved Photo Electron Spectroscopy (ARPES). Notion of a
pseudogap (PG) refers to various systems where a gap $E_{g}$ in their bare
electronic spectra is partly filled showing subgap tails. The best known
examples are the tails in the Density of States (DOS) due to disorder \cite
{lifshits} or the Urbach tails in the subgap optical absorption due to
thermal fluctuations \cite{iosel}. But the PG is especially pronounced in
cases where the bare gap is opened spontaneously as a symmetry breaking
effect which was the subject of detailed experimental studies \cite
{pg-ch,optics,pes,nad-opt,brill}. In quasi-$1D$ conductors it is known as
the Peierls-Fr\"{o}hlich instability leading to the Charge Density Wave
(CDW) formation, as well as to analogous spin-Peierls and Spin Density Wave
states \cite{gruner}. Here the picture of the PG has been suggested
theoretically \cite{lra} in relation to absence of the long range order
(LRO) in $1D$ CDWs. Similarly in High-$T_{c}$ materials the gap opening
mechanisms (still of a disputable origin) which are not stabilized yet by
the LRO are considered. They are supposed to be responsible for the
pseudogap which opens at higher energy scale than the sharp gap appearing
below the transition temperature to the superconducting state \cite{pseudo}.
We shall consider generic $1D$ semiconductors and concentrate on systems
with the dimerized ground state like the well known polyacethylene $(CH)_{x}$
\cite{yu-lu,polymers} or some CDWs with the 2-fold commensurability 
like $NbS_3$ \cite{nad-opt}. Properties of incommensurate CDWs (blue bronzes,
tri- and tetra-chalcogenides of transition metals) \cite{gruner} are further
complicated by interference of the gapless collective mode and we shall
consider them separately.

For $1D$ systems with only a discrete symmetry an absence of the LRO is not
drastically important at low enough temperatures. At finite temperature
there is a remnant concentration $n_{s}\sim \exp [-E_{s}/T]$ of topological
solitons (kinks with the energy $E_{s}$) commuting different domains of the
order parameter $\sim \pm \Delta _{0}$. The midgap electronic states
associated with these solitons create a finite DOS $\sim n_{s}$ at the
former Fermi level $E=0$ and originate related optical features (see \cite
{yu-lu,polymers} for a review and \cite{matv-81} for a systematic theory),
hence there is no true gap at finite $T$. Neglecting this exponentially
freezing contribution, we are left, at first sight, with a sharp electronic
gap $E_{g}^{0}=\Delta _{0}$. But what happens instead is that, even at $T=0$,
 rather smeared edges appear at $\pm \Delta _{0}$ while the spectrum
extends deeply inwards the gap. This effect is particularly pronounced 
in $1D $ because of the edge singularity $\sim (E-E_{g}^{0})^{-1/2}$ at 
$E\approx E_{g}^{0}$ in the bare DOS which smearing is considered to be the
most evident signature of the PG formation below $E_{g}$. In the Peierls
state the PG effect is further enhanced by strong 
electron-phonon ({\it e-ph)} interactions between electrons and quantum
fluctuations of the gap amplitude $\delta =\Delta (x,t)-\Delta _{0}$.
Stationary excitations (eigenstates of the total {\it e-ph }system) are now
the selftrapped states, polarons or solitons, which energies $W_{p}$ or
$W_{s}$ are fractions of  $\Delta _{0}$, see the review \cite{braz-84}. The
states close to the bare electron edge $\Delta _{0}$ can be observed only via
instantaneous measurements like optical or X-ray absorptions or tunneling. The
spectra of these nonstationary states fill the range $\Delta _{0}>E>W_{p}$ for
single electronic (PES) or $2\Delta _{0}>\omega >2W_{s}$ for $e-h$ (optics)
processes. Particularly near $\Delta _{0}$ the states resemble free
electrons in the field of uncorrelated quantum fluctuations of the lattice 
\cite{braz-76}. Here the selftrapping has not enough time to be
developed. But approaching the exact threshold they evolve towards eigenstate
which are selftrapped states accompanied by excitations of their dress. This
picture describes coexistence of the PG region $\Delta _{0}>|E|>W_{p},W_{s}$
and the exact gap $|E|<W_{p},W_{s}$ in a similarity with the High-$T_{c}$
superconductors.

It should be stressed in this respect that there {\em cannot be a common PG}
 for processes characterized by different time scales. We should
distinguish ( as stated in \cite{braz-78} and stressed also in several
reviews, e.g. \cite{braz-84}) between short living states observed in
optical, EPS (and may be tunneling) experiments and long living states
(amplitude solitons, phase solitons) contributing to the spin
susceptibility, NMR relaxation, heat capacitance, conductivity, etc. States
forming the optical PG are created instantaneously: over times which are 
shorter than the inverse phonon frequencies $\tau _{opt}\sim \hbar
/E_{g}<\omega _{ph}^{-1}$ and many orders of magnitude beyond the life times
required for current carriers, and even much longer times for thermodynamic
contributions. It follows then that the analysis of different groups of
experimental data within the same model \cite{ohio} must be reevaluated. 

While experimental techniques of PES and ARPES are still less accurate than
the traditional optics, their very fast progress through the last decade
allows to rely upon a required accuracy already today or in nearest future.
An apparent advantage of the ARPES is its access to the momentum
distributions of spectral densities. Another particular feature of both PES
and its ARPES version in our applications is its decoupling from the final
state Coulomb interactions which affect drastically the optical intergap
absorption. In this respect the PES differs also from the traditional
spectroscopy by internal absorption from a core level; their the final state
interaction between the band particle and the remnant hole, originates a
seminal problem of the X-ray edge singularity in metals \cite{nozieres}.

A theory of the subgap absorption in optics has been developed already: for
a general type of polaronic semiconductors \cite{iosel} with an emphasis to
long range Coulomb effects and for the one dimensional Peierls type system
with an emphasis to solitonic processes \cite{yu-lu,kiv-86} - the last work
is closest to our targets. Here we shall address the PES and ARPES in $1D$
systems. Methodologically all these studies deal with quantum transitions
between distant field configurations which generalize the WKB method for a
single degree of freedom.  These classically forbidden nonlinear processes,
well localized in space and time, are described by extremal trajectories
called the instantons.
The
methods of optimal trajectories in the functional space have been initiated
by the problem of quantum decay of a metastable state \cite{meta} (that is
of a ''false vacuum'' \cite{coleman}) and for quantization of solitons \cite
{dashen} (see the book \cite{raja} for a review). The relevant article on
this line \cite{iord-rash} was devoted to selftrapping barriers for
transformation of a bare electron to the polaron in $3D$ systems.

The plan of our article is the following. In Sec. II we describe a general
scheme for calculating the ARPES probabilities within the adiabatic
approximations, that is based on smallness of typical phonon frequencies
with respect to energies of electron transitions. In Sec. III we explore
applications to particular models. In Sec. IIIA we consider a tutorial zero-
dimensional model for an electron coupled to a single oscillator mode and
show that results of the adiabatic approximation coincide with exact
calculations. In Sec. IIIB we study the transition rate for shallow subgap
states in a $1D$ system, both near the bare gap and at the polaronic
threshold. In Sec. IIIC we consider in details the Peierls model in the
half-filled band case. In Sec. IIID we present the ARPES theory for this
model. Beyond the PES, in Sec. IIIE we also extend earlier calculations \cite
{su-83,kiv-86} for the intragap optical absorption related to creation of
kinks pairs. In Sec. IIIF we consider relations of the general approach to
the model of quantum instantaneous disorder. Finally Sec. IV is devoted to
discussions conclusions.

\section{General relations.}

\subsection{Adiabatic approximation.}

The ARPES \cite{pes} means an absorption of a high energy $\Omega _{0}$
photon taking off an electron from the crystal which is then analyzed in
energy $E_{out}$ and momentum $P_{out}$. Thus one obtains an information on
the spectral density of the hole left beyond. The transition rate $I(\Omega
,P)$, where $\Omega =\Omega _{0}-E_{out}$ and $P=-P_{out}$, is proportional
to the imaginary part of the single-electron retarded Green function

\begin{equation}
I(P,\Omega )\propto Im\int dx e^{-iPx}\int_{0}^{\infty }dTe^{i\Omega
T}G(x,T,0,0).  \label{arpes}
\end{equation}
(Since now on we shall omit all constant factors and take the Plank constant 
$\hbar =1$; $\Omega $ will be measured with respect to a convenient level:
the band edge or the middle of the gap.) The simple PES nonresolved in
momenta measures the integrated absorption intensity $I(\Omega )=\int
dpI(p,\Omega )$.

We shall use the adiabatic (Born-Oppenheimer) approximation. Electrons are
moving in the slowly varying phonon potential, e.g. $Q(x,t)$, so that at any
instance $t$ their energies $E_{j}(t)$ and wave functions $\psi_{j}(x,t)$ 
are defined from the stationary Schroedinger equation for the
instantaneous lattice configuration and they depend on time only
parametrically. The intensity can be written in the form of a functional
integral $D[Q(x,t)]$ over lattice configurations $Q(x,t)$ 
\begin{equation}
 I(\Omega ,P)\propto \int dx e^{-iPx}\int_{0}^{\infty }dT\int D[Q(x,t)]
\Psi_0 (x,T;[Q])\Psi_0^{\ast} (0,0;[Q])\exp (-S[Q])  \label{I}
\end{equation}
This equation is already written in the Euclidean space $it\rightarrow t$
which is adequate for studies of classically forbidden processes \cite
{iosel,raja,iord-rash}.
 Here $\Psi_0 (x,t;[Q])$ is the  wave function of the added $(N+1)$th
particle  in the
instantaneous field $Q(x,t)$ which corresponds to the energy level
  $E_0(t)=E_0[Q(x,t)]$   inside the gap. 
 The  effective action $S=S[Q(x,t)]$ is expressed via
Lagrangians $L_{j}$ as 
\begin{equation}
S={\int_{\infty }^{0}dtL_{0}+\int_{0}^{T}dt(L_{1}-\Omega )+\int_{T}^{\infty
}dtL_{0}}  \label{S}
\end{equation}
where indices $j=1,0$ label systems with and without the additional
particle state (actually the hole). Their typical structure is 
\begin{equation}
L_{j}=\int dx\ \frac{\mu }{2}(\partial _{t}Q)^{2}+V_{j}(Q)
\,,\;\mu =const\sim \omega _{0}^{-2}  \label{L}
\end{equation}
where $\omega _{0}$ is a bare phonon frequency (implying the dispersionless
phonon branch). A nature of the field $Q$, its potentials $V_{j}(Q)$ and the
action upon electrons differ for various models as described in Sec.III.
Usually 
\begin{equation}
V_{j}[Q(x,t)]=\int dx\frac{{Q}^{2}}{2g^{2}}+\epsilon _{j}[Q]\,,\;\mu
g^{2}=\omega _{0}^{-2}  \label{V}
\end{equation}
which contains the bare harmonic term $\sim {Q}^{2}$ and the adiabatic
contribution from the energy of the electron system in the $j$-th excited
state. The electronic terms $j=0,1$ correspond to adiabatic ground states of 
$N$ and $N-1$electrons in the instantaneous field $Q(x)$. For calculations
of subgap processes only lowest localized states $j=0,1$ are relevant, while
other states belong to the continuum spectrum above the gap. 
Then $\Psi $ in
(\ref{I}) is the wave function $\Psi _{0}$ of the split off singly filled
electronic state with the energy level $E_{0}$ inside the gap and $
V_{1}=V_{0}+E_{0}$.

Within the PG region the main contribution to $I$ comes from the saddle
points of the action $S$: $\delta S/\delta Q=0$ $\partial S/\partial T=0$,
the last equation determines the value $T=T(\Omega )$ as 
\begin{equation}
L_{1}(T)-L_{0}(T)=E_0(T) = \Omega.  \label{T}
\end{equation}
(The same relation holds  at the point $t=0$ if one substitute
 $0, T \to  t, t+T$  in Eq. (\ref{S}) and find the minimum over $t$).
We shall find also, in Sec. IIID  devoted to particularities of the ARPES,
circumstances when the extremum must be determined for the whole 
expression
under the integral in the definition (\ref{I}), including the wave functions
in the prefactor of (\ref{I}). Any nonzero contribution requires for a finite
action $S_{0}<\infty $ which selects configurations $Q(x,t)$ deviating from
the ground state only within a finite space-time region. Such  extremal
solutions with finite actions are called the instantons \cite{raja}, their
trajectories correspond to tunneling in the real time. 

\subsection{Translational mode.}

Within the saddle point approximation the transition rate is given by 
\begin{equation}
I(p,\Omega )=I_{0}\exp (-S_{0}),  \label{X1}
\end{equation}
The prefactor $I_{0}=I_{T}I_{X}\dots $ comes from integration over
deviations $\delta Q$, $\delta T$ around the extremal solution. E.g. the 
factor $I_{T}=\left( \partial ^{2}S/\partial T^{2}\right) ^{-1/2} = 
\sqrt{dT/d\Om}$ comes from the
integration over $T$. Usually it can be taken after the extremal is
determined. But there are cases when $I_{0}$ should be determined
selfconsistently, like in statistical physics looking for the minimum of a
free energy, rather then simply of an energy. This option appears naturally
in comparison of PES and ARPES where it comes from treatment of the
translational invariance. Apparently an essential contribution to the action
from the ''quantum entropy'' $\delta S=-\ln I_{0}$ may come only from
integration over particular zero  modes appearing because of
continuous degeneracy. The contribution of usual nondegenerate modes must be
small by definition of a well defined extremum.

The  extremal solution can be written as $Q_{0}(x-X, t)$, $\Psi
_{0}(x-X, t)$ where the collective coordinate $X$ of the instanton
corresponds to the translational invariance which originates one zero mode
in the functional integration around the saddle point solution.
We expand the field $Q(x,t)$ in the vicinity of instanton solution
as
\beq
Q(x,t) = Q_0(x - X(t), t) + \eta(x-X(t), t) 
\eeq
with the orthogonality condition $\int dx  Q_0 \eta =0$, where 
$\eta(x,t)$ contains only nonzero modes.
Integrating over the zero mode is carried out 
 by means of Faddeev-Popov \cite{fad}
method inserting the identity
\beq
1= \int D[X(t)] \delta(\int dx \eta_0(x -X(t),t) Q(x,t)) J,\quad
J = \prod_t \int  dx \partial_x \eta_0(x-X(t),t) Q(x,t),
\eeq
where $\eta_0 = \partial_x Q_0 (x-X(t), t)/\sqrt{\int (\partial_x Q)^2 dx}$ 
is the
normalized zero mode eigenfunction.
The integration over $X(t)$ can be done exactly.
Taking into account only terms containing $X(t)$ we rewrite
(\ref{I}) as
\beq
 I_X \propto \int dx e^{-iPx}\int_{0}^{\infty }\int D[X(t)] J
\Psi_0 (x -X(T),T;[Q])\Psi_0^{\ast} (-X(0),0;[Q])\exp 
\left[- \frac{1}{2}\int dt M(t)
\dot{X}^2(t)\right] ,
\eeq
where $J = \prod_{t_i} \sqrt{M(t_i)}$, and 
  $M(t)$ is the translational effective mass of the
instanton at a given moment $t$:
\[
M(t)= \mu\int dx(\partial _{x}Q)^{2}.
\]
Introducing the integration over $X_1 =X(0)$, $X_2 = X(T)$
one obtains 
\beq
I_X \propto \int \frac{dX_1}{\sqrt{M(0)}} \frac{dX_2}{\sqrt{M(T)}}
 \int dx e^{-iPx}\int_{0}^{\infty }\int D[Q(x,t)]
\Psi_0 (x -X(T),T;[Q])\Psi_0^{\ast} (-X(0),0;[Q])I_1 I_2 I_3,
\eeq
where
\[
I_1 = \int_{X(-\infty)= 0}^{X(0) = X_1} D[X(t)] \prod_{t_i \in (-\infty, 0)}
\sqrt{M(t_i)} e^{-\int_{-\infty}^{0}dt M \dot{X}^2/2 },\,
I_2 = \int_{X(0)=X_1}^{X(T) = X_2} D[X(t)] \prod_{t_i \in (0, T)}
\sqrt{M(t_i)} e^{-\int_{0}^{T} dt M \dot{X}^2/2 },
\]
\[
I_3 = \int_{X(T)=x_2}^{X(\infty) = 0} D[X(t)] \prod_{t_i \in (T, \infty)}
\sqrt{M(t_i)} e^{-\int_{T}^{\infty}dt  M \dot{X}^2/2 }.
\]
Each integral $I_i$ is easy calculated, after the transformation 
$M\dot{X}^2 = \dot{Y}^2$,  with the help of the result \cite{dashen}
\[
\int_{x(0)=x_1}^{x(T) = x_2} D[x(t)] e^{-\int_0^T (\dot{x}^2 + V(x))dt} =
\left|\frac{\partial^2 S_0}{\partial x_1 \partial x_2}\right|^{1/2}
e^{-S_0}, 
\]
where $S_0$ is the saddle point action.
 After  simple integrations  over $dx$, $dX_1$, $dX_2$ we arrive at 
\beq
I_X \propto \mid\Psi_P (T)\mid^2 \exp\left[{-\int_{0}^{T}dt\frac{P^{2}}{2M(t)}}\right],
\label{L-P}
\end{equation}
where $\Psi_p$ is the Fourier transform of the wave function $\Psi_0$.
 
Contrary to stationary solutions, the solitons (polarons, kinks), here $M(t) 
$ depends on time along the instanton trajectory $Q_0(x,t)$.
 The exponential term in  (\ref{L-P}) gives the  contribution to the total 
action  which is important 
 at very large $P$ because generally it is of the order of $M^{-1}\sim \omega
_{0}^{2}$. It provides an important, sometimes leading effect upon the
edge shape: it is singular along the instanton term with $M$ vanishing at
large $|t|$ when the space localization is week and $M^{-1}(t)$ diverges.

\subsection{Zero dimensional reduction.}

 As a rule the space-time differential nonlinear equations for extremals are
not solvable, hence one is bound to variational procedures.
We  shall use the following ansatz which is actually an one-parameter reduction
 of the functional space. The instanton trajectory 
must satisfy the condition (\ref{T}) which means that  there is an
electron level $E_0 = \Om$  at times
$t=0,T$. It seems reasonable that the trajectory with only one
local level inside the gap will 
be a good approximation for the extremal action.
There is also a special advantage 
that  for models considered below we know   exact
 solutions of the stationary
problem  $Q_0 (x,a)$ depending on some continuous parameter $a$ with
the local level $E_0 (a)$ inside the gap.
Therefore we shall search the instanton  (time-dependent) solution in
the form $Q_0(x, a(t))$. After  integration around
saddle point    $Da(t)$  we find  finally for ARPES intensity
\beq
I(P, \Om) \propto \sqrt{\frac{dT}{d\Om}} \mid \Psi_p (a_T)\mid^2
e^{-S_0 },
\label{IPO}
\eeq
where $S_0$ is the instanton action which is extremum  over $T, a(t)$
of
\beq
S= \int_{-\infty}^{\infty} dt [ f(a) \dot{a}^2 +\frac{P^{2}}{2M(a(t))}
+ V(a(t)],
\label{sarp}
\eeq
where 
\beq
f(a)=\frac{\mu }{2}\int
dx\left( \frac{\partial Q_{a}(x,t)}{\partial a(t)}\right) ^{2}
\eeq
is the variable effective mass,
$V(a(t)) = V_1 (a(t)) - \Om $, for $ \in [0, T]$, or $V(a(t)) = V_0(a(t))$
 otherwise.

In some cases the form-factor $\Psi_P^2$ gives  
the main contribution to the function  $I( P, \Om)$. 
Then we must find the 
extremal solution for a modified action 
$S \to S- \log \mid \Psi_P \mid^2$. Such a possibility
will be considered in details in Sec. IIID.     

Since  the PES intensity is obtained simply by integration  of ARPES one,
we obtain
\beq
I(\Om)  \propto \left(\int \frac{dp}{2\pi}e^{-p^2 l^2 /4}
\mid \psi_p (T)\mid^2 \right)
\sqrt{\frac{dT}{d\Omega}} \exp[-S_0],
\label{pre0}
\eeq
where 
\[
l^{2} = \int_0^T \frac{dt}{M(a(t))},
\]
and $S_0$ is the extremum of the action (\ref{sarp}) without 
the kinetic term $P^2/2M$.
The form-factor $F = \int dP \mid \psi_p (T)\mid^2\exp(-P^2l^2/2)$ in
(\ref{pre0})  provides a kind of Debye-Waller reduction of intensity. Thus for
small  $l$ in compare to the localization length $\xi $ of $\Psi $ (short times
near the free edge) $F\approx 1$ while for large $l$ $\gg \xi $ (long
times $T$ near the absolute threshold) $F\sim |\Psi _{P=0}(T)|^{2}\sim
\xi /l$ with $\xi $ being the polaronic width. In our examples typically it
will hold $l\ll \xi $.

For  typical potentials  $V(a)$ (see Fig. 1 - 4 below)
the extremum solution  with a finite action exists
only in some region of $\Om$ where
the potential curve $V_1(a) - \Om$ (or $V_1(a) -\Om + P^2/2M(a)$ for
the case of ARPES) crosses both the curve $V_0(a)$ and $V=0$.
It takes place if the minimum of the potential $V_1 -\Om$
 ( or $V_1 -\Om + P^2/2M$) is placed below the minimum of the 
potential $V_0$  (min $V_0 =0$).
If   $a_T$ is the point where
\beq 
V_0(a_T) = V_1(a_T) -\Om \quad {\rm (PES )}, \qquad 
V_0(a_T) = V_1(a_T) -\Om + P^2/2M(a_T) \quad {\rm (ARPES )},
\label{VaT}
\eeq
and $a_f$ is the solution of
\beq
V_1 (a_f) - \Om =0 \quad {\rm (PES )}, \qquad 
 V_1(a_f) -\Om + P^2/2M(a_f)=0 \quad {\rm (ARPES )},
\label{Vaf}
\eeq
then the extremal solution is the closed
trajectory, described by the equation  
\beq
f(a) \dot{a}^2 = \tilde{V}(a),
\label{ii}
\eeq
where $\tilde{V} =  V_0(a)$, $a \in [0, a_T]$, and
$ \tilde{V} =  V_1(a) - \Om +P^2/2M(a)\}$, $a \in (a_T, a_f]$   
for ARPES,
or the same expression without the kinetic potential $P^2/2M$ for PES.
The Eq. (\ref{VaT}) is a consequence of the Eq. (\ref{T}), it implies
a continuity of the velocity $\dot{a}$  at  the point $a_T$. 
The  Eq. (\ref{Vaf}) reads that the velocity $\dot{a} =0$ at the point $a_f$.

The instanton  equation  (\ref{ii})
describes the motion of some particle of zero 
energy
with variable mass $2f(a)$ in the inverted potential $-\tilde{V}$.
The  trajectory starts at $t= -\infty$ from the point $a=0$,
reaches the points $a_T$ at $t=0$, $a_f$ at $t =T/2$, after which the 
particle moves back across the point $a_T$ at  $t=T$ to the
initial point $a=0$ at $t \to \infty$.
The instanton action (for ARPES) can be expressed as 
\begin{equation}
S_0 =4\int_{0}^{q_{T}}dq\sqrt{f V_{0}}+4\int_{q_{T}}^{q_{f}}dq\sqrt{
f(V_{1}-\Omega +P^2/2M(a) )}.
\label{actin}
\end{equation}

\section{Results for absorption intensities.}

\subsection{Zero-dimensional case}

We shall start with a tutorial
 example where the general approximate scheme
can be compared with exact calculations. Consider a particle interacting
with a single quantum oscillator at zero temperature. Physically it can be a
problem of the Jahn-Teller center or a zero order approximation for the
small radius polaron. The total Lagrangian is 
\begin{equation}
L=\frac{m}{2}\dot{q}^{2}-\frac{m}{2}\omega _{0}^{2}q^{2}-gqN,
\end{equation}
where $q$ is the oscillator coordinate, $m$ is its mass, $g$ is the
interaction constant, and ${N}$ is the electron's occupation number. The
potentials $V_{0}$ and $V_{1}$ 
\[
V_{0}=\frac{m\omega _{0}^{2}}{2}q^{2},\qquad 
V_{1}=\frac{m\omega _{0}^{2}}{2}
q^{2}-gq=\frac{m\omega _{0}^{2}}{2}(q-q_{0})^{2}-W_{0}\,;\;
q_{0}=\frac{g}
{m\omega _{0}^{2}}\,,\;W_{0}=\frac{g^{2}}{2m\omega _{0}^{2}}. 
\]
are shown in Fig. 1.
In compare to $N=0$, for $N=1$ the oscillator states $n$ are shifted by $
q_{0}$ in coordinate and by $-W_{0}$ in energy. Adiabatically allowed
transitions take place only at the given $q$ that is they explore the region 
$q=0$, $\Omega =0$. Taking into account the quantum character of the 
coordinate 
$q$, the transitions are found to the lower energies down to $W_{0}$. The
region $0>\Omega >-W_{0}$ 
corresponds to pseudogaps in more complex systems.

\begin{figure}[tbph]
\begin{center}
\includegraphics[width=3.0in]{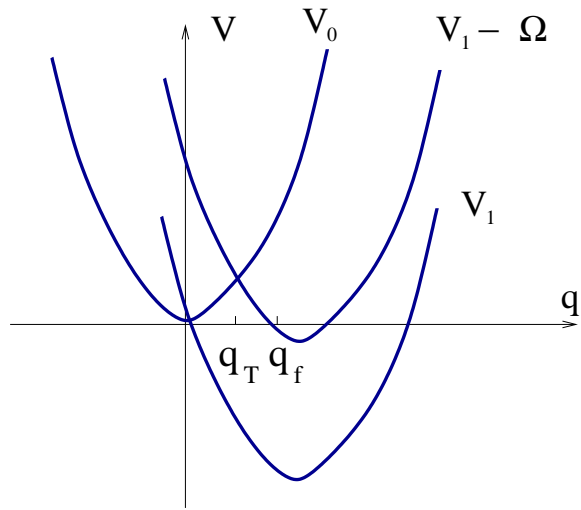}
\caption{ Potentials $V_{0}$, $V_{1}-\Omega $. }
\end{center}
\end{figure}
The solutions of (\ref{VaT}), (\ref{Vaf}) read
\[
 q_{T}=-\Omega /g, \qquad q_{f}=2g/(m\omega _{0}^{2}).
\]
The instanton action is easy calculated and 
for the absorption rate  we arrive at
\begin{equation}
I(\Omega )\propto \frac{1}{\sqrt{\Omega + W_0}}
\exp (-\frac{W_{0}}{\omega _{0}})\exp\left[ \frac{\Omega
+W_{0}}{\omega _{0}}\log \frac{ e W_{0}}{(\Omega +W_{0})}\right] .  
\label{Pd=0}
\end{equation}
Expanding the exponent near the extremum point $\Omega =0$ we obtain
\begin{equation}
I(\Omega )\propto \exp \left[ -\frac{\Omega ^{2}}{2W_{0}\omega _{0}}\right] .
\label{o1}
\end{equation}
which describes smearing of the adiabatically allowed electronic edge. 
This is the Gaussian with the width 
$\delta \Omega =\sqrt{W_{0}\omega _{0}}\gg \omega
_{0}$ which justifies the adiabatic approximation. At the lower boundary $
\Omega =-W_{0}$ of the exact spectrum the probability has a finite
exponentially small value which is approached with an infinite slope.

For this simple example $W(\Omega)$ can be calculated
exactly in the representations of eigenstates. Indeed,
\begin{equation}
I(\Omega ) = \sum_{n}|\langle n,1|\Psi ^{+}|0,0\rangle
|^{2}\delta (\Omega -E_{n}+E_{0}),
\end{equation}
where $|0,0\rangle $ is the ground-state of the system without electron with
the energy $E_{0}=\hbar \omega _{0}/2$, and $|n,1\rangle $ is the $n$-th
excited state ($n=0,1,...$) with one electron and the energy
$E_{n}=\hbar \omega _{0}(n+1/2)-W_{0}$.
The wave functions are the ones for the harmonic oscillator which are
centered at $q=0$ for $|0,0\rangle $ and at $q=q_{0}$ for $|n,1\rangle $. In
the limit $\omega_{0}\rightarrow 0$ and $n=(\Omega +W_{0})/\omega_{0}\gg 1$
we find
\begin{equation}
I(\Omega )=
\frac{\exp [^{-(\delta y)^{2}/2}]}{\sqrt{2\pi n}\omega_{0}}\exp 
\left[ n\log {\frac{e(\delta y)^{2}}{2n}}\right],
   \label{Posc}
\end{equation}
where $ \delta y = g/\sqrt{m \omega_0^3}$.
This equation  reproduces the  result (\ref{Pd=0}).

\subsection{Shallow states in a 1D system.}

Consider the electron (hole) states in a 1D dielectric near an edge of a
conducting (valence) band. We shall take into account the dispersionless
phonon field $Q(x,t)$ with the bare frequency $\omega _{0}$ which interacts
locally with the electron via the deformation potential with a coupling $g$.
Within the adiabatic approximation the action has the form (again at
imaginary time) 
\begin{equation}
S=\int dx\left\{ \int_{-\infty }^{+\infty }dt\left( \frac{{1}}{2\omega
_{0}^{2}}\left( \frac{\partial Q}{\partial t}\right) ^{2}+\frac{1}{2}Q^{2}\right) 
+\int_{0}^{T}dt\left( \frac{1}{2m}\left| \frac{\partial \Psi }{\partial
x}\right| ^{2}+gQ\Psi ^{\dagger }\Psi -\Omega \right) \right\} , \label{h2}
\end{equation}
It is well known that the stationary ($\dot{Q}=0$) extremum of $S$
corresponds to the selftrapped state (the polaron) \cite{rashba}. We arrive
directly at Eqs. \ref{L},\ref{V} with $E$ and $\Psi $ as solutions of the
Schr\"{o}dinger equation: 
\begin{equation}
-\frac{1}{2m}\frac{d^{2}\Psi }{dx^{2}}+gQ\Psi =E\Psi
\,,\;E=E(t)=E[Q(x,t]_{t}.  \label{se}
\end{equation}
The minimum of the initial potential $\min V_{0}=0$ is achieved at the
uniform configuration $Q=0$. The minimum of $V_{1}$ takes place at the
polaron state $\min V_{1}=W_{p}=-W_{0}=-mg^{2}/24$. The minimum of 
$V_{1}-\Omega $ must be lower than the minimum of $V_{0}$, hence 
$-W_{0}<\Omega <0$.

Since the exact solution of the complete time dependent extremal equations
is not known, we shall follow a variational approach. We will search for the
instanton solution in the form which simulates the stationary solution but
with the variable time dependent parameter $B(t)$ which we chose as the
inverse localization length: 
\begin{equation}
Q(x,t)=-\frac{B^{2}(t)}{mg}\frac{1}{\cosh ^{2}B(t)x},\qquad \Psi =
\sqrt{\frac{B(t)}{2}}\frac{1}{\cosh B(t)x},\qquad E(t)=-\frac{B^{2}}{2m}.
\end{equation}
The stationary solution corresponds to $B(t)=B_{0}=mg^{2}/2$. In terms of 
$b(t)=B(t)/B_{0}$ the Lagrangians become 
\begin{equation}
L_{0}=W_{0}\left[ \frac{C_{0}}{\omega _{0}^{2}}b\dot{b}^{2}+2b^{3}\right]
\,,\;L_{1}=W_{0}\left[ \frac{C_{0}}
{\omega _{0}^{2}}b\dot{b}^{2}+2b^{3}-3b^{2}\right] ,  \label{ss}
\end{equation}
where $C_{0}=4+2\pi ^{2}/15\approx 5.3$. The instanton equation describes
the motion of a  zero energy
particle, with a coordinate $b$ and with a variable mass 
$f\sim b$, in the inverted potentials 
\[
V_{0}=W_{0}2b^{3}\;{\rm and}\;V_{1}-\Omega =W_{0}\left[ 2b^{3}-3b^{2}\right]
-\Omega . 
\]
which are shown in Fig. 2.
\begin{figure}[tbph]
\begin{center}
\includegraphics[width=3.0in]{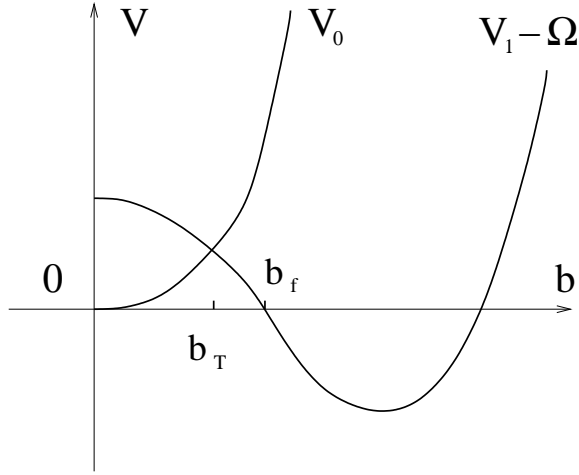}
\caption{ Potentials $V_{0}$, $V_{1}-\Omega $. }
\label{}
\end{center}
\end{figure}
 The value $b_{T}=\sqrt{\left| {\Omega }/{E_{0}}
\right| }$,  the
value $b_{f}$ at the turning point is determined by the condition
$V_1 (b_f) - \Om =0$.

For the PES the action is 
\[
S_{0}\approx 4\sqrt{C_{0}}\frac{W_{0}}{\omega _{0}}
\left[
\int_{0}^{b_{T}}\sqrt{2}b^{2}db+
\int_{b_{T}}^{b_{f}}\sqrt{(2b^{3}-3b^{2}-\frac{\Omega }{W_{0} })b}\,db\right] 
\]
and we arrive at the following results.

Near the free electronic edge $\Omega =0$ we have 
\begin{equation}
I(\Omega )\propto \mid \Omega \mid ^{-1/4}\exp \left[ -\frac{8}{9}\sqrt{
\frac{C_{0}}{6}}\frac{(-\Omega )^{3/2}}{\omega _{0}W_{0}^{1/2}}\right]
\label{om=0}
\end{equation}
The characteristic width of the edge is $\delta \Omega \sim (\omega
_{0}^{2}W_{0})^{1/3} \gg \omega _{0}$ 
which justifies the adiabatic
approximation.

In the vicinity of the absolute edge $\Omega \approx -W_{0}$ we obtain 
\begin{equation}
I(\Omega )\propto \frac{1}{\sqrt{\Omega +W_{0}}}\exp [-2\frac{W_{0}}{\omega
_{0}}]\exp \left[ 2\sqrt{\frac{C_{0}}{3}}\frac{\Omega +W_{0}}{\omega _{0}}
\ln \frac{e(\sqrt{3}-1)^{2}W_{0}}{(\Omega +W_{0})}\right] .  \label{om=wp}
\end{equation}
In this limit the probability is finite, exponentially small in the
adiabatic parameter $W_{0}/\omega _{0}$ and it shows a weak singularity in
its derivative. \bigskip The prefactors in above formulas come from $
I_{T}\sim (d^{2}S/d\Omega ^{2})^{1/2}$. 
The form-factor is $F\approx 1$
according to Sec. II. Indeed, in spite of a formal divergence of $l$, within
limits of the adiabatic approximation ($|\Omega |\gg \omega _{0}$ or 
$(\Omega +W_{0})\gg \omega _{0}$) it stays below the 
localization length $\xi$. 
Thus $l/\xi \sim (\omega _{0}/\Omega )^{1/2}\ll 1$ for small $\Omega $
and $l/\xi \sim (\omega _{0}\ln [W_{0}/(\Omega +W_{0})])^{1/2}$ for 
$\Omega \approx -W_{0}$.

Peculiarities of the ARPES will be studied in the next section within a
richer Peierls model.

\subsection{The Peierls model}

We consider now the PES absorption spectrum in the gap region for a
half-filled Peierls model. In the ground state the electron spectrum has the
form $E^{2}=v_{F}^{2}p^{2}+\Delta _{0}^{2}$ and since now on we shall put
the Fermi velocity $v_{F}=1$. The gap $2\Delta _{0}\sim \epsilon _{F}\exp
[-1/\lambda ]$ is opened as a result of symmetry breaking lattice
deformations. We shall consider the case of a dimerization
(trans-Polyacethylene, Spin-Peierls systems) which is described by the model
with the real order parameter $\Delta $ taking equilibrium values $\pm
\Delta _{0}$. The excited states are solitons (kinks), polarons and
bisolitons (kink-antikink pairs) which are characterized by electron levels
localized deeply within the gap (see the review \cite{braz-84}). The
adiabatic approximation is valid when the electron transition energies are
much larger than the phonon frequency $\omega _{0}\ll |E_{i}-E_{j}|$. For
characteristic $|E_{i}-E_{j}|\sim \Delta _{0}$ this constraint coincides
with the applicability condition for the Peierls model in general \cite
{braz-76}. Away from the ground state the field $\Delta (x,t)$ extends a
role of the field $Q$ from the previous section. The first difference is
that $\Delta \neq 0$ already in the ground state. The second one is that the
whole sea of electrons at $E<-\Delta _{0}$ contributes to the selftrapping
energy \cite{braz-84} and to the instanton terms.

The effective Euclidean action $S$ consists of the kinetic and the bare
potential lattice energy and of the sum over filled electron levels $%
E_{\alpha }$. 
\[
S\{\Delta (x,t)\}=\int dxL_{j}dt,\quad L_{j}=\int dx\frac{\dot{\Delta}^{2}}{%
\pi \lambda \omega _{ph}^{2}}+V_{j}[\Delta (x,t)]. 
\]
where $\omega _{ph}=\sqrt{\lambda }\omega _{0}$ is the bare frequency of the
phonon responsible for the dimerization while $\omega _{0}$ is the amplitude
mode frequency in the dimerized state $\Delta \neq 0$. The index $j$
characterizes perturbations in the set $\{\alpha _{j}\}$ of filled
electronic states in the field $\Delta $. The potential energy in the $j$-th
electronic state is 
\[
V_{j}[\Delta (x,t)]=\int dx\,\frac{\Delta ^{2}}{\pi \lambda }+\sum_{\alpha
_{j}}E_{\alpha _{j}}\{\Delta (x,t)\}-W_{gs}, 
\]
which we have defined relative to the ground state energy $W_{gs}$ at $%
\Delta (x,t)=\Delta _{0}$ for a system of $N=N_{a}$ electrons ($N_{a}$ is
the number of sites in the chain).

The PES absorption adds one particle (hole) to the system, then the polaron
state is formed by local deformations of the field $\Delta (x)$ which
originate the pair of split off electronic levels $\pm E_{0}$ localized
deeply in the gap. In optical absorption\cite{kiv-86} an electron is excited
across the gap, then a pair of distant kinks is formed with electron level
placed exactly in the center of the gap ($E_{0}=0$). The crossover solution
describing the states with just one pair of localized electronic levels $\pm
E_{0}$ is known \cite{braz-81,lanl}: 
\begin{equation}
\Delta _{s}(x)=\Delta _{0}(1-\tanh a[\tanh (\Delta _{0}x\tanh a+\frac{a}{2}%
)-\tanh (\Delta _{0}x\tanh a-\frac{a}{2})]).  \label{bs}
\end{equation}
Depending on the parameter $a$ it describes evolution from the shallow
polaron at $a\rightarrow 0$ to the pair of kinks at $a\rightarrow \infty $.
In the first case the parameter $a\ll 1$ becomes equivalent to $b$ from the
previous section. In the second case $a\gg 1$ becomes a distance between
divergent kinks. For the configuration \ref{bs} the potential energy
functional $V$ and the energy of the local level $E_{0}$ are given by

\begin{equation}
V_{\nu }(a)=\nu E_{0}+\frac{4}{\pi }\sqrt{\Delta _{0}^{2}-E_{0}^{2}}-\frac{4%
}{\pi }E_{0}\cos ^{-1}\frac{E_{0}}{\Delta _{0}}\,,\;E_{0}=\frac{\Delta _{0}}{%
\cosh a},  \label{vbs}
\end{equation}
where $\nu =0,1,2$ is the number of particles (electrons and holes) added to
the system.
The wave function of the intragap state can be written as \cite{matv} $\Psi
_{0}(x)\propto \sqrt{\Delta _{0}^{2}-\Delta _{s}(x)^{2}}$.

The case $\nu =0$ describes the state which evolves from the ground one
without perturbing the occupation numbers: the level $-E_0$ is doubly filled
while the level $E_0$ is empty. Naturally the minimum of $V_{0}$ corresponds
to the uniform configuration $\Delta (x)=\Delta _{0}$ at $a=0$ without a
split-off level. The function $V_{0}(a)$ monotonically increases from $%
V_{0}(0)=0$ to $V_{0}(\infty )=2W_{s}$, ( $W_{s}=2/\pi \Delta _{0}$ is the
total soliton energy) with the asymptotic behavior. 
\begin{equation}
V_{0}(a)\approx \frac{4}{3\pi }\Delta _{0}a^{3},\quad a\rightarrow 0,\quad
V_{0}(a)\approx 2W_{s}-2E_0(a),\quad a\rightarrow \infty.  \label{as}
\end{equation}

For the polaron $\nu =1$ (either one electron at $E_{0}$ or one hole at $%
-E_{0}$) the equilibrium state corresponding to the minimum of $V$ is
achieved at $a=a_{0}$ where 
\[
\sinh a_{0}=1\,,\,\min {V}_{1}{=V(a}_{0})=W_{p}=2^{3/2}\Delta _{0}/\pi
\,,\,E_{0}=\Delta _{0}/\sqrt{2}. 
\]
The limiting values of $V_{1}(a)$ are $V_{1}(0)=\Delta _{0}$, $V_{1}(\infty
)=2W_{s}$.

The case $\nu =2$ corresponds to either an exciton (one electron and one
whole at $\pm E$ that is each of these levels is singly occupied) or to
electron or hole bipolarons (all states $\pm E$ are either filled or empty).
The excitonic state plays a principal role in the subgap optical absorption
problem \cite{kiv-86,su-83}. The equilibrium state is achieved at $a=\infty $
where $V=2W_{s}$ and $E_{0}=0$. The function $V_{2}(a)$ decreases
monotonically from $V_{2}(0)=2\Delta _{0}$ to $V_{2}(\infty )=2W_{s}$. The
dependencies $V_{\nu }(a)$ are shown in Fig. 3.

\begin{figure}[tbph]
\begin{center}
\includegraphics[width=3.0in]{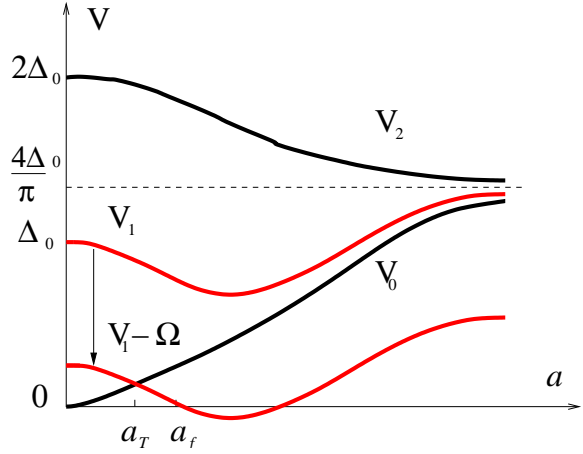}
\caption{Plots of $V_{\protect\nu }(a)$ for $\protect\nu =0,1,2$.}
\label{}
\end{center}
\end{figure}

The above $a-$ dependent family of solutions is the only one which provides
exactly one pair of discrete levels in the gap. Its perturbations originate
only shallow levels located near the edges $\pm \Delta _{0}$ which are not
important for the described processes. Therefore we will choose the
configuration $\Delta (x,t)=\Delta _{s}(x,a(t))$ of (\ref{bs}) taken with
the time dependent parameter $a$. The kinetic term in the Lagrangian
becomes $f(a)\dot{a}^{2}$ with 
\begin{equation}
f(a(t))=\frac{1}{\pi \omega _{0}^{2}}\int dx\left( \frac{\partial \Delta _{s}
}{\partial a}\right) ^{2}.
\end{equation}
The function $f(a)$ increases from $f(0)=0$ reaching the maximum value $
f\approx 1.55\Delta _{0}/(\pi \lambda \omega _{0}^{2})$ at $a\approx 0.78$,
and then decreases with asymptotics 
\begin{equation}
f(a)\approx \frac{2}{3}\frac{C_{0}}{\pi }\frac{\Delta _{0}a}{\omega _{0}^{2}}
,\;a\rightarrow 0;\qquad f(a_{0})\approx \frac{1.53\,\Delta _{0}}{\pi \omega
_{0}^{2}};\qquad f(\infty )=\frac{2}{3}\frac{\Delta _{0}}{\pi \omega _{0}^{2}
}.  \label{f}
\end{equation}
with the same coefficient $C_{0}$ as in the previous section.

The translational mass becomes 
\begin{equation}
M(a)=\frac{2}{\pi \omega _{0}^{2}}\int dx\left( \frac{\partial \Delta }{
\partial x}\right) ^{2}=\frac{8\Delta _{0}^{3}}{g^{2}\omega _{0}^{2}}\left[ 
\frac{\tan ^{3}a}{3}-\frac{a\cosh a-\sinh a}{\cosh ^{3}a}\right] .
\end{equation}
The function $M(a)$ monotonically increases with asymptotics 
\begin{equation}
M(a)\approx C_{M}\frac{\Delta _{0}^{3}}{\omega _{0}^{2}}a^{5} ,\quad
C_M =\frac{32}{15}, \quad
a\rightarrow 0; \qquad
M(a_{0})\approx 0.49\frac{\Delta
_{0}^{3}}{\pi \omega _{0}^{2}};\qquad M(\infty )=\frac{16}{3\pi }
 \frac{\Delta _{0}^{3}}{\omega _{0}^{2}}.  \label{M}
\end{equation}

Following the prescriptions of Sec. II,
 we can describe now the PG region $
W_{p}<\Omega <\Delta _{0}$. 
In the limiting cases the results are
qualitatively similar to the ones for the shallow polaron.

Near the free edge, $\Omega \approx \Delta _{0}$ we find 
\begin{equation}
I(\Omega )\sim (\Delta _{0}-\Omega )^{-1/4}\exp \left[ -\frac{32\sqrt{C_{0}}%
}{9\pi }\frac{(\Delta _{0}-\Omega )^{3/2}}{\omega _{0}\sqrt{\Delta _{0}}}%
\right] .  \label{om-del}
\end{equation}
with the same coefficient $C_{0}$ as in (\ref{ss}).

Near the polaronic energy $\Omega \approx W_{p}$ we find 
\begin{equation}
I(\Omega )=const\frac{1}{\sqrt{\Omega -W_{p}}}\exp \left[ -C_{1}\frac{\Delta
_{0}}{\omega _{0}}\right] \exp \left[ C_{2}\frac{(\Omega -W_{p})}{\omega _{0}%
}\log \frac{C_{3}\Delta _{0}}{(\Omega -W_{p})}\right]  \label{pom}
\end{equation}
Here the numerical coefficients $C_{1} = 0.34$, $C_{2}= 5.2$,
$C_{3} = 0.1$.

\subsection{ARPES intensities.}

For  case of ARPES the potential $V_1$ acquires an additional term
$P^2/2M$, therefore the closed trajectory is allowed in the region 
$(\Omega ,P)$ where (See Fig. 4) 
\begin{equation}
\tilde{V}_1 (a_m) =
V_{1}(a_{m})+P^{2}/2M(a_{m})-\Omega < 0\,,\; d\tilde{V}_1 /da 
\vert_{a=a_m} =0
\label{reg1}
\end{equation}
Here the value $a_{m}$ is the point of a minimum of the potential 
$\tilde{V}_{1}(a)$, 
which exists for any value of the momentum $P$. In the limit $P=0$
we have $a_{m}=a_{0}$. For other $\Omega $, $P$ there are no solutions with
a finite action so that the transition rate is zero. In contrast to the PES
case, the closed trajectory exists for any frequency $\Omega >W_{p}$ in the
region of the momentum $P$ satisfying the condition (\ref{reg1}).

\begin{figure}[tbph]
\begin{center}
\includegraphics[width=3.0in]{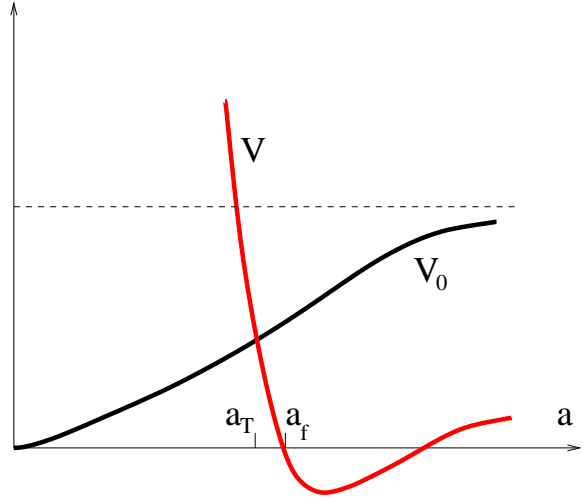}
\caption{ Plots of $V_{0 }(a)$  and ${V_1} +P^2/2M(a) - \Om $.}
\end{center}
\end{figure}

Consider  most important cases where analytical solution can be found.

A. In a vicinity of the absolute edge $\Omega \approx W_{p}$ the absorption
appears at $\Omega \geq V_{1}(a_{m})+P^{2}/2M(a_{m}) \equiv \tilde{W}_p$.
 Here, near the
polaron edge, the translational mass is almost constant staying near $
M(a_{m})$ hence the momentum appears simply through the shift $\Omega
\Rightarrow \Omega -P^{2}/2M(a_{m})$.
\beq 
I (P, \Om )  \propto \frac{1}{\sqrt{\Omega - \tilde{W}_p}} \exp[-S_0].
\eeq
with
\begin{equation}
S_0 = \frac{\tilde C_{1}}{\sqrt{\pi \lambda }}
\frac{\Delta_{0}}{\omega _{0}} 
+ \left[\tilde C_{2}\frac{(\Omega -\tilde{W}_{p})}{\sqrt{\pi\lambda }\omega _{0}}
\log \frac{\tilde C_{3}\Delta _{0}}{(\Omega-\tilde{W}_{p})}\right].
\label{pom1}
\end{equation}
The
result for the absorption looks like the Eq. (\ref{pom}) for the PES with $
W_{p}\Rightarrow W_{p}(P)$ but the coefficients $C_{j}$ become functions of $%
P$: $\tilde{C}_{1},\tilde{C}_{2},\tilde{C}_{3}$. They are defined as
\[
\tilde C_{1}=4\sqrt{\pi \lambda }\frac{\omega _{0}}{\Delta _{0}}
\left[\int_{0}^{a_{T}}da\sqrt{f(a)V_{0}(a)}+
\int_{a_{T}}^{a_{f}}da\sqrt{f(a)(\tilde{V}_{1}-W_{p})}\right],
\]

\[
\tilde C_{2}=\sqrt{\pi \lambda }\omega _{0}
\frac{2^{3/2}\sqrt{f(a_{m})}}{\sqrt{\tilde{V}_{1}^{\prime \prime}(a_{m})}}
\,,\;
\tilde C_{3}=\frac{e}{2}(a_{f}- a_{T})^{2}
\frac{\tilde{V}_{1}^{\prime \prime }(a_{m})}{\Delta _{0}}.
\]
and at $P\rightarrow 0$ they coincide with the numerical ones $C_{1}$, $
C_{2} $, $C_{3}$ defined above in Sec. IIIC.

B. A vicinity of the free edge $\Omega \approx \Delta _{0}$ 
 is very particular for the
ARPES. It is dominated by shallow fluctuations determined by 
small $a \ll 1$ where the
kinetic energy diverges as $P^{2}/M\sim a^{-5}$ according to
 (\ref{M}). 
The form-factor  $\mid \Psi_P\mid^2$ acquires an explicit form 
\begin{equation}
\mid \Psi _{P}\mid ^{2}\approx \frac{\pi }{\Delta _{0}a_{T}}\cosh ^{-2}{
\frac{\pi P}{2\Delta _{0}a_{T}}}  \label{FF}
\end{equation}
To simplify appearance of cumbersome relations we shall use below
dimensionless units $\Delta _{0}=1$ for the momentum $P$ and 
energies which
also will be counted with respect to $\Delta _{0}$.
In the limit $a_t, \, a_f \ll 1$ equations (\ref{VaT}), (\ref{Vaf})
are reduced to
\beq
-\eps -\frac{a_T^2}{2} + \frac{\omega_0^2 P^2}{C_M a_T^5} = 0,
\label{eps1}
\eeq
\beq
-\eps + \frac{4}{3\pi}a_f^3 - \frac{a_f^2}{2} + 
\frac{\omega_0^2 P^2}{C_M a_f^5} = 0,
\label{eps2}
\eeq
where $\eps = (\Om - \Delta_0 )/\Delta_0$ and the term 
$-a_T^2/2$ comes from the expansion of the local level energy $E_0$.
 A simple analysis
of  (\ref{eps1}), (\ref{eps2}) shows that $a_f - a_T = o(a_T)$,
and  first term gives the main contribution to the action (\ref{actin}),
 so that 
\beq
S_0 =  \frac{8 \sqrt{2C_0}}{9\pi} \frac{\Delta_0 }{\omega_0} 
a^3_T .
\label{arps}
\eeq
The general constraint is $S_0\gg 1$ hence $a_T\gg \omega _{0}^{1/3}$. It
assures us also in the condition $M\gg \Delta _{0}$ (the mass $M$ must stay
above the free band edge mass) which requires for a weaker inequality 
$a_T\gg \omega $.

We have  different regions where one of terms of Eq. (\ref{eps1})
is small in comparison with others.

B1.  Consider at first the region lying deeply enough within the PG:
\[ (\om_0 P)^{4/7} \ll -\eps \ll 1 
\]
Here the characteristic values of $a_T$ are not too small so that
one can neglect the kinetic energy $\sim P^{2}\omega _{0}^{2}/a_T^{5}$ in
compare to the energy $ -a_T^{2}/2$ of the localized level: 
$P^{2}\omega _{0}^{2}/a_T^{5}\ll a_T^{2}$. 
Then $a_T\approx \sqrt{-2\eps }$,  $
S\sim |\eps |^{3/2}/\omega _{0}$ and $d\Om /dT \sim a_T$. The
momentum dependence comes only from the form-factor (\ref{FF}) 
but it can be
still appreciable: $P/a_T$ is limited by a big quantity ($P/a_T\ll \eps
^{5/4}/\omega _{0}$ $\gg \omega _{0}^{-1/6}$) that is allowed to be large.

The total intensity is 

\begin{equation}
I(\eps ,P)\propto \frac{1}{\mid \eps\mid ^{3/4}}\exp 
\left[-\frac{32 \sqrt{C_0}}{9\pi \omega_0 }\mid\eps\mid^{3/2} - \frac{\pi P}
{\mid 2 \eps\mid^{1/2}}\,\right]  \label{I-B1}
\end{equation} 
In the limit $-\eps \ll  \sqrt{P\omega_0}$ 
 the contribution from $\mid\Psi_p\mid^2$ 
(the second term in the exponent) dominates, 
while in the opposite case 
the main
contribution comes from  the term $S_0$. 
The line of the maximum intensity 
$-\eps \sim \sqrt{\omega_0 P}$
will show up in
the ARPES plots as a {\em quasi spectrum} which intensity decreases
exponentially  with growing momentum: 
$I(P)\sim P^{-3/8}\exp [-const P^{3/4}\omega _{0}^{-1/4} ] $.

B2. Consider a marginal frequency lying very close to the free edge $\eps
=0$ for both signs of $\eps $: 
\[
\omega _{0}\ll |\eps |\ll \left( \omega _{0}P\right) ^{4/7}\ll 1\;{\rm 
for\;}P\gg \omega _{0}^{2/5}. 
\]
In this case the instanton kinetic energy and the binding of the electron
almost compensate each other: $P^{2}/2M\approx a_T^2 /2 \gg |\eps |$
which gives us $a_T=\left( (\omega _{0}P)^{2}/C_{M}\right) ^{1/7}$ and we
obtain 
\begin{equation}
\,I(\eps, P)\propto \frac{1}{ P^{3/7}}\exp \left[ -\pi 
\frac{\pi (C_{M})^{1/7} }{\omega
_{0}^{2/7}} P^{5/7}   -  
\frac{8\sqrt{2C_0}}{9\pi (C_M)^{3/7} \omega_0^{1/7}} P^{6/7}
\right] \label{I-B2} \end{equation}
The form-factor $\Psi_P^2$  gives the main contribution (the first term)
 to the exponent of (\ref{I-B2}) . 

B3. Consider now a new region of positive frequencies $\eps >0$:
\[
 \eps \gg (\omega _{0}P)^{4/7}$ but $\eps \ll \omega _{0}^{1/3}P^{2}.
\]
 The first inequality ensures that the excess
frequency is absorbed mainly by the kinetic energy with a negligible 
($ a_T^2/2 \ll \eps $) contribution from the electronic state which is
still localized. The second inequality supports the adiabatic
condition $S_0\gg 1$. We find 
\[
a_T\approx \frac{ (\omega
_{0}P)^{2/5}}{(2C_{M}\eps )^{1/5}}, \quad \frac{dT}{d\Om} \propto
\frac{P^{6/5}}{\eps^{13/5}}.
\]
The ARPES intensity becomes
\begin{equation}
I(\eps, P) \propto \frac{P^{1/5}}{\eps^{11/10}}\exp \left[ -
\frac{\pi (2C_{M})^{1/5}}{\omega_0^{2/5}}
P^{3/5}\eps^{1/5}-
\frac{8\sqrt{2C_0} \omega_0^{1/5}}{9\pi (2C_M)^{3/5}} 
\frac{P^{6/5}}{\eps^{3/5}} \right]  
\label{B3} 
\end{equation}
The form-factor contribution, the first term in the exponent, always
dominates but the second term is also big thus contributing to the
dependence $I(\eps ,P)$. 
The total expression in the exponent of (\ref{B3})
looks non monotonous similar to the case B1 but now the minimal is not
physical: its position $\eps \sim (\omega _{0}P)^{3/4}$ would fall too
low, into the region B2. Contours of constant intensity are close to the
descending line $\eps \sim \omega _{0}^{2}/P^{3}$. The contribution 
(\ref{B3}) from the {\em  fast moving instantons} can be observable only
below the intensive free electronic absorption at $\eps \approx P^{2}/2$
that is at $ P\gg \omega _{0}^{2/5}$ (a similar constraint already appeared
for the case B2).

For the ARPES intensities we must, in principle, minimize the total 
effective action with  the prefactor term $\mid\Psi_P\mid^2$,  since
in some cases  this term gives a main contribution to the
exponent. Then the action is changed as
\[
S \to S - \log \mid\Psi_P\mid^2 
\]
The extremal solution, as before, satisfies the Eq. (\ref{ii}).
The  turning point $a_f$ is not changed and is defined by the 
Eq. (\ref{Vaf}), but the equation defining the point $a_T$
becomes more complicated. Instead of the Eq. (\ref{VaT})
we obtain
\beq
V_0 (a_T) - \tilde{V}_1 (a_T) = \frac{1}{2} 
\frac{\partial \log \mid\Psi_P \mid}
{\partial a} \left( \sqrt{\frac{V_0}{f}} + \sqrt{\frac{\tilde{V}_1}{f}}\right)
\Biggl|_{a=a_T}.
\label{VaT1}
\eeq 
This equation leads to the discontinuity of the potential $\tilde{V}$
at the point $a_T$ and, due to the equation of motion (\ref{ii}),
leads  also to the discontinuity of the "velocity" $\dot{a}$.  
The expression 
for the ARPES intensity has the same form (\ref{ii})
 with the action (\ref{actin}). The only difference is the shift
of the point $a_T$ due to Eq. (\ref{VaT1}). 

For a considered in B1-B3 regions the point $a_T$ is defined
by (\ref{eps1}) with the additional term $\sim P\omega_0 /a_T$.
The simple analysis  shows that 
this term  is small in comparison
with the kinetic part $P^2 /2M$ or with the  term  $a^2/2$,
therefore it can be neglected. Thus,   
 the main contribution to the action is not changed,
and  all above results are retained.    
  
\subsection{Optical absorption: effects of confinement.}

The same method is applicable to the subgap optical absorption problem which
was basically studied already in \cite{yu-lu,kiv-86}. We shall see that more
details can be easily obtained. In this case we must use the potential $%
V_{2} $, defined in (\ref{vbs}), instead of $V_{1}$. The term $V_{2}$
describes the first excited state when the optical photon, assisted by a
lattice quantum fluctuation, creates an $e-h$ pair with levels $\pm E(t)$
spanning the whole interval $|E|<\Delta _{0}$. The transition rate is given
by the modified Eq. (\ref{X1}). For the vicinity of the free electronic edge $%
\Omega \sim 2\Delta _{0}$ we easily find, in analogy with (\ref{om-del}),
that

\begin{equation}
I(\Omega )\propto \exp \left[ -\frac{16\sqrt{2C_{0}}}{9\pi }\frac{(2\Delta
_{0}-\Omega )^{3/2}}{\omega _{0}\sqrt{\Delta _{0}}}\right] .
\end{equation}
This law was noticed in \cite{kiv-86}, while with a different coefficient in
the exponent. Actually it was obtained already in \cite{braz-76} by another
method which was later reproduced in \cite{ohio}. Near the absolute
absorption edge $\Omega \approx 2W_{s}=4\Delta _{0}/\pi $ we obtain 
\begin{equation}
I(\Omega )\propto \exp [6\sqrt{f(\infty )}\sqrt{\Omega -2W_{s}}]  \label{opt}
\end{equation}
where $f(\infty )$ is given by Eq. (\ref{f}). Near this edge the absorption
is due to tunneling to the final state formed by the two diverging kinks
with the distance parameter $a\rightarrow \infty $. (This regime was studied
for the first time in \cite{kiv-86} but the above analytical asymptotics was
not found.) The law (\ref{opt}) is different from (\ref{pom}) which was
typical for all polaronic thresholds. The particularity is related to a very
fast, $\sim \exp [-a]$, decrease of interactions for diverging solitons
which determine the threshold. In terms of the effective particles the
difference is that near the polaronic threshold the turning point approaches
the potential extremum while for the solitonic threshold it climbs
asymptotically to the plateau (the dashed line at the Fig.3).

While the law (\ref{pom}) is very robust, e.g. the same as for shallow
polarons, the solitonic threshold is extremely sensitive to perturbations.
The most drastic effect comes from the confinement energy $\delta V_{\nu
}=Fa $ with the constant confinement force $F$ originated by lifting of the
ground state degeneracy. The confinement is always produced by the
interchain coupling \cite{braz-80}. But even for a single chain it appears
in cases of a build-in alternation of unit cells which interferes with the
spontaneous dimerization like in $cis-(CH) {x}$ \cite{braz-81,matv}. In
principle, the solitonic terms for this problem have been found exactly in
the frame of the microscopic model of the ''combined Peierls state'' \cite
{braz-81,matv} and they can be used in our calculations for an arbitrary
strength of the force $F=\gamma \Delta _{0}/\xi _{0}$. But here we shall
consider only a weak confinement $\gamma \ll 1$ which preserves the local
structure of solitons but prevents their divergence at large $a$. This
effect can be taken into account by adding the term $\gamma \Delta _{0}a$ to
the potentials $V_{0}$, $V_{2}$ taken in zero order in $\gamma \ll 1$. (For
the PES this effect is not relevant since the minimum of the polaronic
potential $V_{1}$ is achieved at finite value $a_{0}$ where confinement is
not important yet.) As a result the minimum of the potential $V_{2}$ is
shifted to the finite point $a_{\gamma }\approx \log (16/\pi \gamma )/2$,
where $V_{2}^{\prime \prime }(a_{\gamma })=2\gamma $. The optical absorption
edge is shifted up to $W_{\gamma}=4\Delta _{0}/\pi +\gamma \Delta _{0}\log
(16/\pi \gamma )/2$ and the transition rate $I(\Omega )$ is given by Eq. (\ref
{pom}) if substitute $W_{p}\rightarrow W_{\gamma }$, $V {1}\rightarrow V {2}$%
, $a_{0}\rightarrow a_{\gamma }$, $a_{1}\rightarrow \cosh ^{-1}(\pi /2)$.

The problem of solitonic absorption, without the confinement, was considered
in \cite{su-83,kiv-86} by means of a different quasi-classical approaches.
Our results supply the absent analytical asymptotics in this case and
provide a new description for the general case of confinement. Further
complications are coming from the long range Coulomb interactions which must
be studied before comparison with existing experiments.

\subsection{Quantum fluctuations as an instantaneous disorder.}

In a one dimensional system the X-ray or optical absorption near the band
edge can be viewed as the one in a system with a quenched disorder which is
emulated by instantaneous quantum fluctuations. This conclusion was achieved 
\cite{braz-76} by analysis of the perturbation series for the polarization
operator which is not the most efficient way to do. Later applications of
this idea did not add any prove \cite{ohio}. Here we shall derive this
result within the approach of functional integrals which will provide a
model independent limit for over approximate analysis of more complicated
regimes and systems.

Consider a particle interacting with a media of harmonic oscillators. This
model embraces the shallow polaron case of Sec.B and the limit of $%
\Omega\approx\Delta_0$ of Sec.D. The probability is given by the action

\begin{eqnarray}
S\{Q,\Psi ;T) &=&\int d^{D}x\int_{-\infty }^{+\infty }dt \left( \frac{{1}}{%
2\omega _{0}^{2}} \left( \frac{\partial Q}{\partial t}\right) ^{2}+ \frac{1}{%
2}Q^{2}\right) \\
&&+ \int d^{D}x\int_{0}^{T}dt\left( \frac{1}{2m} \left| \frac{\partial \Psi 
}{\partial x}\right| ^{2}+gQ\Psi ^{\dagger }\Psi \right).
\end{eqnarray}
We can integrate out the field $Q$ at all $x$ and $t$ to arrive at the
action which is defined only at the interval $(0,T)$, c.f. \cite{iosel}: 
\[
S\{\Psi ;T)= \int d^{D}x\left[ \int_{0}^{T}dt\left( \frac{1}{2m} \left| 
\frac{\partial \Psi }{\partial x}\right| ^{2}\right) - \frac{1}{4}%
g^{2}\omega_{0}\int_{0}^{T}dt_{1} \int_{0}^{T}dt_{2}\rho (x,t_{1})\rho
(x,t_{2}) \exp[-\omega _{0}|t_{1}-t_{2}|]\right] 
\]
where $\rho =|\Psi |^{2}$. The result will prove that the characteristic
energy scale $\delta _{D}$ is much larger than $\omega _{0}$ which takes
place only at $D<2$. Then the whole time interval is short $T\omega
_{0}\sim\omega _{0}/\delta _{D}\ll 1$ and we can neglect the retardation: $%
\exp[-\omega _{0}|t_{1}-t_{2}|]\rightarrow 1$. Now the last term in $S\{\Psi
;T)$ can be decoupled back by the Hubbard-Stratonovich transformation via
the {\em time independent } auxiliary field $\zeta$ to give us 
\[
S\{\Psi ,\zeta ;T)= \int d^{D}x\left[ \int_{0}^{T}dt\left( \frac{1}{2m}
\left|\frac{\partial \Psi }{\partial x}\right| ^{2}\right) +
\zeta(x)\int_{0}^{T}dt_{1}\rho (x,t)+ \frac{1}{g^{2}\omega _{0}}\zeta ^{2}(x)%
\right] 
\]
Finally we can integrate over $\Psi $ to arrive at the transition probability

\[
I(\Omega)\sim \int D[\zeta (x)]\int dT\exp \left[ -\frac{1}{g^2 {\omega _{0}}%
}\zeta ^{2}(x)+T(E[\zeta (x)]-\Omega )\right] 
\]
After rotation to the real time it becomes 
\[
\int D[\zeta (x)\delta (E[\zeta (x)]-\Omega )\exp \left[ -\frac{1}{%
g^{2}\omega _{0}}\zeta ^{2}(x)\right] 
\]
where $E[\zeta (x)]$ is the eigenfunction in the random field $\zeta$ 
\[
\frac{-1}{2m}\left( \frac{\partial }{\partial x}\right) ^{2}\Psi +\zeta
\Psi=E\Psi 
\]
Clearly the dispersion $\delta_0 = g\sqrt{\omega _{0}/2}$ of the field $%
\zeta $ is just the mean square of quantum fluctuations of the phonon field $%
gQ(x,t)$. The well known results for the DOS of disordered systems \cite
{lifshits,disorder} confirm our direct calculations for the time dependent
processes in the pseudogap region and extend them to the nonexponential part
of the spectrum of $\Omega >0$. For $D=0$ ($m=\infty$) and $D=1$ we arrive
correspondingly at the widths $\delta_{0}\sim g\sqrt{\omega _{0}}$ and $%
\delta _{1}\sim (g^{2}\omega _{0})^{2/3}$ to find that in both cases the
condition $\delta _{D}\gg \omega _{0}$ is satisfied at low enough $\omega
_{0}$. (This condition is not satisfied at higher dimensions hence at $D\geq
2$ the quantum fluctuations are not reduced to the random static field but
rather become resolved phonon assistant processes.\cite{braz-78}) An
additional peculiarity of the Peierls model is that, in microscopic units of 
$\Delta _{0}$ and $\xi _{0}$ the coupling constant $g=1$, then the band edge
smearing is given by $\delta _{1}=(\Delta _{0}\omega _{0}^{2})^{1/3}$.

\section{Discussion and Conclusions}

Electronic properties of quasi one - dimensional conductors in the Peierls
state are quite peculiar in several respects. In general it is due to a
strong interaction of CDW deformations with normal electrons which leads to
their fast self trapping. The stationary excited states of the model are
solitons (kinks) and polarons with energies $E_{s},E_{p}<\Delta _{0}$. The
processes related to these nonlinear excitations determine the true gaps $%
2E_{s},2E_{p}$ placed within the pseudogap $2\Delta _{0}$. The single
particle gap, as measured in absorption or tunneling, is opened by polarons
which should exist also in $1D$ systems without symmetry breaking, like the
majority of conjugated polymers \cite{braz-81,matv}. The effect is very
common because in $1D$ semiconductors the selftrapping of free electrons
takes place for any type of e-ph interactions \cite{rashba} while in higher
dimensions the long range interactions are required. Here the minimal role
of the Peierls effect, or at least its partial contribution, is to ensure a
presence of the strong e-ph coupling. The optical threshold exists at $2E_{p}
$ in general. But for systems with degenerate ground states like the
polyacethylene there is also a lower gap at $2E_{s}$ because the lowest
excitations are now topological solitons (kinks) \cite{braz-78,ssh,tlm}. A
weak interchain coupling \cite{les-arcs} preserves the polaronic effects,
while creating a shallow barrier with respect to the selftrapping, thus
allowing for metastable free electronic states and corresponding
transitions. The $3D$ effects on solitonic transitions are more drastic
because of the confinement effect \cite{braz-80,les-arcs}.

Presented results recover the stretch exponential dependencies near the PG
edge (for both PES and optics) 
\begin{equation}
I\sim \exp \left[ -\left( \frac{\Omega }{\delta _{D}}\right) ^{\nu _{D}}%
\right] \,,\;\delta _{D}\sim \omega _{0}^{1/\nu _{D}}  \label{edge}
\end{equation}
with different powers $\nu _{D}>1$.  Near the absolute threshold at $\delta
\Omega =\Omega -W_{p}\ll W_{p}$ we find the same (for the PES) law

\begin{equation}
I\sim \exp \left[ -C\frac{W_{p}}{\omega _{0}}-C^{\prime }\frac{\delta \Omega 
}{\omega _{0}}\ln \frac{W_{p}}{\delta \Omega }\right]   \label{om-ln-om}
\end{equation}
which differ only by numerical coefficients among different models. The
first constant term in the exponent describes the adiabatic reduction for
the probability of creation of the stationary polaron at $W_{p}$. The second
term gives the law for approaching this threshold. At first sight, this
quasi linear dependence of the exponent $\sim \frac{\delta \Omega }{\omega
_{0}}\ln \frac{W_{p}}{\delta \Omega }$ is weaker than the one $\sim \ln 
\frac{W_{p}}{\delta \Omega }$ coming from the pre-exponential factors $\sim
(\delta \Omega )^{-n}$ but it is not true: by definition we are working at $%
\delta \Omega $ which can be small only in compare to $W_{p}$ but must be
large in compare to $\omega _{0}$. Hence $\frac{\delta \Omega }{\omega _{0}}%
\ln \frac{W_{p}}{\delta \Omega }\gg \ln \frac{W_{p}}{\delta \Omega }$ and
the exponential term determines the profile near the absolute threshold.
The same law holds for the two particle process (the internal optical
absorption) if the ground state is nondegenerate. But for a system with a
spontaneous symmetry breaking like the Peierls model for the polyacethylene
the threshold dependence of the exponent changes from $\sim \delta \Omega
\ln \delta \Omega $ to a stronger one $\sim \sqrt{\Omega -2W_{s}}$ with the
threshold $2W_{s}$ being the energy of a solitonic pair.

The momentum dependence of the intensity $I(\Omega ,P)$ as recovered by the
ARPES shows a rich variety of regimes. Only near the absolute threshold the
law (\ref{om-ln-om}) is simply generalized by adding the polaron kinetic
energy: $W_{p}\Rightarrow W_{p}+P^{2}/2M_{p}$. But the region near the free
electronic threshold $\Omega \approx E_{g}$ demonstrates several nontrivial
regimes. The differences come from two effects achievable for shallow states.
One is the momentum dependence of the form factor - the momentum
distribution of the split off state. Another one is the inertial part of the
action coming from a drastic dependence of the kinetic mass on the
localization parameter which evolves in time along the instanton process.
One of unexpected results is that at large $P$ the law (\ref{edge}) changes
to a non monotonous function of $\Omega $ which would show itself as a quasi
spectrum (a line of maxima of $I(\Omega ,P)$) within the pseudogap. Another
result is an appearance of optimal localized fluctuations at an elevated
kinetic energy so that they show up above the PG region at $\Omega >0$.
These observations may be more general than the studied one dimensional
adiabatic models.

In conclusion, we have calculated intensities of subgap photo electronic
transitions by means of a functional integration over the lattice
oscillation modes. By this virtue we have studied the problem of
photoemission from the pseudogap region for typical one-dimensional models.
We have found general expressions for the transition rate and studied the
asymptotic behaviors near absorption edges, both below the free electronic
edge and approaching the lower true one. It was shown that the main
contributions to the transition rates comes from instanton configurations of
the phonon field. The Peierls model was investigated in particular details
in relation with its numerous applications. Peculiarities of the ARPES
regime are coming from unexpected effects of the instanton motion.

For the intragap optical absorption problem we have calculated the
asymptotic behavior of the absorption coefficient near the threshold for
creation of pairs of solitons and studied effects of confinement.

\acknowledgments
S. M. acknowledges the hospitality of the Laboratoire de Physique Theorique
et des Modele Statistiques, Orsay and the support of the CNRS and the ENS -
Landau foundation.

\end{document}